\documentclass[a4paper,fleqn,usenatbib]{mnras}
\pdfoutput=1

\usepackage[T1]{fontenc}
\usepackage{ae,aecompl}

\usepackage{graphicx}	
\usepackage{amsmath}	
\usepackage{amssymb}	
\usepackage{pdflscape}

\title[Intracluster Age Gradients]{Intracluster Age Gradients In Numerous Young Stellar Clusters}

\author[K. V. Getman et al.]{
K. V. Getman,$^{1}$\thanks{E-mail: kug1@psu.edu (KVG)}
E. D. Feigelson,$^{1}$
M. A. Kuhn,$^{2,3}$\newauthor
M. R. Bate,$^{4}$
P. S. Broos,$^{1}$
G. P. Garmire$^{5}$
\\
$^{1}$Department of Astronomy \& Astrophysics, 525 Davey Laboratory, Pennsylvania State University, University Park PA 16802\\
$^{2}$Instituto de Fisica y Astronomia, Universidad de Valparaiso, Gran Bretana 1111, Playa Ancha, Valparaiso, Chile\\
$^{3}$Millenium Institute of Astrophysics, Av. Vicuna Mackenna 4860, 782-0436 Macul, Santiago, Chile\\
$^{4}$Department of Physics and Astronomy, University of Exeter, Stocker Road, Exeter, Devon EX4 4QL, UK\\
$^{5}$Huntingdon Institute for X-ray Astronomy, LLC, 10677 Franks Road, Huntingdon, PA 16652, USA
}

\date{Accepted for publication in MNRAS, 2018 January 31. Published in MNRAS, 476, 1213G.}

\pubyear{2018}

\begin{document}
\label{firstpage}
\pagerange{\pageref{firstpage}--\pageref{lastpage}}
\maketitle

\begin{abstract}
The pace and pattern of star formation leading to rich young stellar clusters is quite uncertain. In this context, we analyze the spatial distribution of ages within 19 young (median $t \la 3$~Myr on the Siess et al. (2000) timescale), morphologically simple, isolated, and relatively rich stellar clusters. Our analysis is based on young stellar object (YSO) samples from the MYStIX and SFiNCs surveys, and a new estimator of pre-main sequence (PMS) stellar ages, $Age_{JX}$, derived from X-ray and near-infrared photometric data. Median cluster ages are computed within four annular subregions of the clusters. We confirm and extend the earlier result of Getman et al. (2014): 80\% percent of the clusters show age trends where stars in cluster cores are younger than in outer regions. Our cluster stacking analyses establish the existence of an age gradient to high statistical significance in several ways. Time scales vary with the choice of PMS evolutionary model; the inferred median age gradient across the studied clusters ranges from 0.75~Myr~pc$^{-1}$ to 1.5~Myr~pc$^{-1}$. The empirical finding reported in the present study -- late or continuing formation of stars in the cores of star clusters with older stars dispersed in the outer regions -- has a strong foundation with other observational studies and with the astrophysical models like the global hierarchical collapse model of V{\'a}zquez-Semadeni et al. (2017).
\end{abstract}




\section{Introduction} \label{sec_introduction}

While it is recognized that most stars in the Galaxy form today in clusters with $10^{2}-10^{4}$ stars \citep{Lada2003}, the pace and pattern of star formation leading to rich stellar clusters is quite uncertain.  It is unclear whether clusters form rapidly in a burst of star formation of a large and dense molecular core \citep{Elmegreen2000}, or over millions of years in filamentary concentrations of a turbulent molecular cloud \citep{Krumholz2007}.  

Observationally, comparison of  individual cluster members with evolutionary tracks on a Hertzsprung-Russell diagram (HRD) typically show a wide age dispersion, but this is difficult to interpret due to various observational and theoretical reasons such as photometric variability, multiplicity, and accretion history \citep{Preibisch2012}.  The situation is confused by the presence of mass segregation in some clusters which may or may not arise from a delayed formation process for massive OB stars \citep{Zinnecker2007}.  These issues are debated even for the most well-studied case where the full stellar population is identified, the Orion Nebula Cluster \citep[ONC;][]{Hillenbrand1997,Huff2006,ODell2009,Reggiani2011,Jeffries2011,DaRio2014, Messina2017, Beccari2017}. 

In this context, \citet*{Getman2014b} reported empirical evidence for a spatial gradient of stellar ages within two young rich stellar clusters in the nearest giant molecular cloud, the ONC ionizing the Orion Nebula and NGC 2024 ionizing the Flame Nebula.  The result is that stars in cluster cores appear younger (that is, formed later) than stars in cluster peripheries.  These findings are independently supported by spatial gradients in the disk fractions of these clusters, and are not related to mass segregation as the analysis is restricted to stars with masses between $\sim 0.2$ and 1.2~M$_\odot$. The result is based on a new estimator of pre-main sequence (PMS) stellar ages  derived from X-ray and near-infrared photometry, $Age_{JX}$, that avoids some of the pitfalls of ages derived from HRDs \citep{Getman2014a}. 

The observed spatial age gradient is not predicted by simple cluster formation models; a centrally concentrated cloud core would show stars in the core are older, not younger, than in the periphery \citep{PP2013}.  More complex models are needed involving processes of dynamical expansion of older sub-clusters, sub-cluster mergers, and/or resupply of core gas by in-falling filaments \citep{Getman2014b}.  Other studies independently indicate stars in the dense Trapezium core are younger than more dispersed stars in the Orion Nebula Cluster \citep{ODell2009,Reggiani2011,Beccari2017}.

The purpose of the present study is to extend the search for stellar age gradients from the two Orion cloud clusters in \citet{Getman2014b} to a larger sample of 19 clusters in more distant molecular clouds. Our effort is based on the datasets from two recent X-ray/infrared stellar cluster surveys: Massive Young Star-Forming Complex Study in Infrared and X-ray \citep[MYStIX;][]{Feigelson2013} and Star Formation in Nearby Clouds \citep[SFiNCs;][]{Getman2017a}.  MYStIX provides cluster membership catalogs for 20 OB-dominated star forming regions (SFRs) at distances from 0.4 to 3.6 kpc \citep{Broos2013}. SFiNCs extends the MYStIX effort to a study of 22 generally nearer SFRs where the stellar clusters are often dominated by a single massive star -- typically a late-O or early-B -- rather than by numerous O stars in the MYStIX fields. The SFiNCs YSO and cluster catalogs are available in \citet{Getman2017a,Getman2017b}. The spatial distribution of stars in these SFRs is complex, and cluster memberships are identified using a maximum likelihood mixture model \citep{Kuhn2014,Getman2017b}.  


The result of our search is that core-halo age gradients are present in the vast majority of the analyzed clusters: delayed or continuing star formation in cluster cores is a general phenomenon.  The evidence requires that large samples of stars be averaged together because individual stellar $Age_{JX}$ estimates are not accurate.   The paper is organized as follows. The age analysis methodology and the cluster samples are reviewed in Section \ref{sec_samples}. The discovered intra-cluster age gradients are presented in Section \ref{sec_results}, and the implications for cluster formation are discussed in Section \ref{sec_discussion}.

\section{Samples And Methods} \label{sec_samples}
\subsection{Cluster Samples} \label{sec_samples_only}
Among the 141 MYStIX \citep{Kuhn2014} and 52 SFiNCs \citep{Getman2017b} clusters that are identified by an objective likelihood-based statistical procedure \citep{Kuhn2014} we select the ones that have: 1) a simple morphology (a single dense core surrounded by a sparser halo); 2) are relatively isolated (without or with only weak secondary sub-clusters present in their vicinity); 3) and have a relatively high number of stars with available age estimates. Nineteen clusters, with 2487 $Age_{JX}$ stars, satisfy these criteria.
\begin{figure}
\centering
\includegraphics[angle=0.,width=100mm]{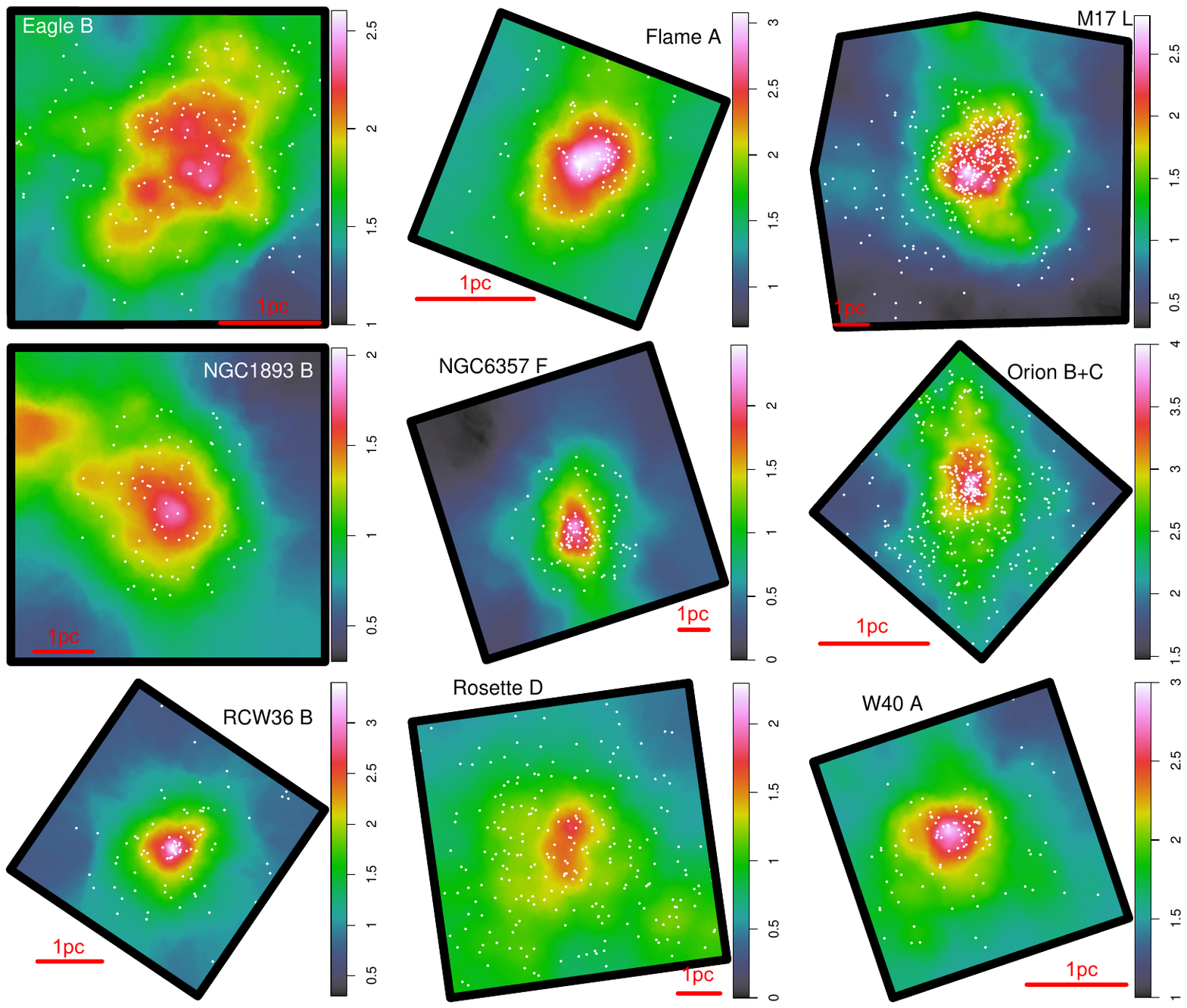}
\caption{Adaptively smoothed projected stellar surface densities of the clusters with a color-bar in units of observed stars per pc$^{2}$ (on a logarithmic scale). Six clusters are shown here; similar panels for the full sample of 19 clusters are presented in the Supplementary Materials.   The white circles mark the positions of stars with available $Age_{JX}$ estimates. The black polygons outline either the original full {\it Chandra}-ACIS-I fields of view for a single dominant cluster or the cutouts of the {\it Chandra} fields to separate the cluster of interest from other nearby clusters. \label{fig_maps_individual}}
\end{figure}

Table \ref{tbl_clusters} lists the cluster's designation and center celestial position taken from \citet[][MYStIX]{Kuhn2014} and \citet[][SFiNCs]{Getman2017b} as well as the distance from the Sun taken from \citet[][MYStIX]{Feigelson2013} and \citet[][SFiNCs]{Getman2017a}. In the case of M~17 and Rosette, the multiple clumpy stellar structures around the primary ionizing clusters identified by Kuhn et al. are combined here into single entities. In the case of Orion, RCW~36, and Be~59 the core and halo structures identified by Kuhn et al. and Getman et al. are treated here as single entities.  
\begin{table}\small
\centering
 \begin{minipage}{80mm}
 \caption{Rich, Isolated Young Star Clusters. Column 1: Star forming region. Column 2: Cluster. The cluster's designation is from \citet[][MYStIX]{Kuhn2014} and \citet[][SFiNCs]{Getman2017b}. Columns 3-4. Celestial coordinates (J2000) for the cluster center, taken from \citet[][MYStIX]{Kuhn2014} and \citet[][SFiNCs]{Getman2017b}. Column 5. Distance from the Sun. The distance values are taken from \citet[][MYStIX]{Feigelson2013} and \citet[][SFiNCs]{Getman2017a}.}
 \label{tbl_clusters}
 \begin{tabular}{@{ \vline }c@{ \vline }c@{ \vline }c@{ \vline }c@{ \vline }c@{ \vline }}
\cline{1-5}
&&&&\\ 
Region & Cluster & R.A. &
Decl. & Distance\\
&&(deg)&(deg)&(pc)\\
(1)&(2)&(3)&(4)&(5)\\
\cline{1-5}
&&&&\\ 
Eagle Nebula & B &   274.675830 &   -13.784167 & 1750\\
Flame Nebula & A &    85.427083 &    -1.903806 &  414\\
M 17 & L &   275.124580 &   -16.179444 & 2000\\
NGC 1893 & B &    80.694167 &    33.421389 & 3600\\
NGC 6357 & F &   261.509170 &   -34.278333 & 1700\\
Orion Nebula & B$+$C &    83.820000 &    -5.389556 &  414\\
RCW 36 & B &   134.863750 &   -43.757222 &  700\\
Rosette Nebula& D &    97.980833 &     4.944167 & 1330\\
W 40 & A &   277.861250 &    -2.094167 &  500\\
Be 59 & B &     0.562134 &    67.418764 &  900\\
Cep A & A &   344.073598 &    62.030905 &  700\\
Cep OB3b (Cep B) & A &   343.446066 &    62.596289 &  700\\
Cep C & A &   346.441016 &    62.502827 &  700\\
GGD 12-15 & A &    92.710405 &    -6.194814 &  830\\
IC 348 & B &    56.141167 &    32.158825 &  300\\
IC 5146 & B &   328.381131 &    47.265152 &  800\\
LkHa 101 & A &    67.542392 &    35.268025 &  510\\
NGC 7160 & A &   328.443151 &    62.585448 &  870\\
SerpensMain & B &   277.492003 &     1.216660 &  415\\
&&&&\\
\cline{1-5} 
\end{tabular}
\end{minipage}
\end{table}

Fig.~\ref{fig_maps_individual} shows the morphology of six clusters; and the full set of panels for the 19 clusters is available in the Supplementary Material. These stellar spatial distributions are presented in the form of an adaptively smoothed projected stellar surface density. The YSO catalogs of \citet{Kuhn2014} and \citet{Getman2017b} serve as the basis for constructing these maps. The maps are constructed using the Voronoi tessellation technique\footnote{Description of the Voronoi tessellation can be found on-line at \url{https://en.wikipedia.org/wiki/Voronoi\_diagram}.} and are presented in units of observed stars per pc$^{2}$ (on a logarithmic scale). These adaptive estimates of the surface density are computed using the {\it R} function {\it adaptive.density} from the {\it spatstat} package \citep*{Baddeley2015}. The celestial positions of the surface density peaks are given in Table \ref{tbl_clusters}.

The purpose of the maps is to emphasize the presence of relatively compact and dense stellar cores (white and/or red) surrounded by extended and shallower halos (green). From the core toward periphery, the stellar surface density typically falls by a factor of $\ga 10$. The clusters are captured by the MYStIX/SFiNCs surveys at the spatial scales of two to several parsecs.

\subsection{Age Methods} \label{sec_ages_only}
The stellar ages for the young stars in the MYStIX and SFiNCs sub-clusters were estimated in \citet{Getman2014a,Getman2017a} using the $Age_{JX}$ method of \citet{Getman2014a}. $Age_{JX}$ is applicable only to low-mass PMS stars ($M<1.2$~M$_{\odot}$ assuming the \citet*{Siess2000} age scale) with reliable measurements of the intrinsic X-ray luminosity and near-infrared $JHK_s$ photometry. The method is based on an empirical X-ray luminosity - mass relation calibrated to well-studied Taurus PMS stars \citep{Telleschi2007} and to theoretical evolutionary tracks calculated by \citet{Siess2000}. X-ray luminosities specify stellar masses; $J$-band luminosities and mass estimates track with PMS evolutionary models, providing stellar ages. While individual $Age_{JX}$ values may be noisy, median ages for stellar clusters and sub-regions have sufficiently low statistical errors allowing discoveries of spatio-age gradients across the MYStIX star forming regions and within the ONC and NGC 2024 clusters \citep{Getman2014a,Getman2014b}. As most standard PMS age methods, $Age_{JX}$ may not provide reliable absolute ages. All sub-region ages derived in this work are relative but are referenced to a uniform time scale.

Time scales change with changing an underlying PMS evolutionary model. Here we consider three different PMS models: the relatively old model of \citet{Siess2000}[hereafter Siess00], the new non-magnetic model of \citet{Dotter2016,Choi2016}[hereafter MIST], and the new magnetic model of \citet*{Feiden2015,Feiden2016}[hereafter Feiden16M]. Age transformations among those models can be obtained by placing numerous artificial young stars on the $(L_{bol},T_{eff})$ diagram, within the parameter ranges of the SFiNCs/MYStIX $Age_{JX}$ stars (see figure in \citet{Richert2017}).

The time scale transformations can be approximated by the following linear functions: $t_{MIST} = -0.45 + 0.83 \times t_{Siess00}$ and $t_{Feiden16M} = -0.31 + 1.67 \times t_{Siess00}$. Compared to Siess00, a newer generation of standard (non-magnetic) evolutionary models, such as MIST, with improved micro-physics (including updated solar abundance scale, linelists, atmospheric convection parameters), predicts systematically younger PMS ages. The MIST time scale is very similar to that of \citet{Baraffe2015} (not shown here in a figure). In contrast, the new generation of magnetic models \citep{Feiden2015,Feiden2016} anticipates systematically older ages. \citet[][and references therein]{Feiden2015,Feiden2016} argue for inclusion of magnetic fields, threaded into the stellar interior, in order to mitigate the ``radius inflation'' effect, that is an under-prediction of stellar radii and over-prediction of effective temperatures in low-mass ($M<1$~M$_{\odot}$) PMS companions of eclipsing binaries \citep{Kraus2015} by standard PMS models.   
\begin{table}\small
\centering
 \begin{minipage}{80mm}
 \caption{$Age_{JX}$ Stars. All MYStIX and SFiNCs stars with available age estimates across the 19 clusters of interest. This table is available in its entirety (2487 stars) in the machine-readable form in the online journal. A portion is shown here for guidance regarding its form and content. Column 1: Star forming region. Column 2: Source's IAU designation. This unique source's name keyword can be used to retrieve other numerous X-ray, near-IR, and mid-IR source's properties from the YSO catalog datasets of \citet[][MYStIX]{Broos2013} and \citet[][SFiNCs]{Getman2017a}. Columns 3-4. Source's celestial coordinates in decimal degrees (J2000). Column 5. Stellar age estimated using the $Age_{JX}$ method of \citet{Getman2014a}. These age estimates are based on the PMS evolutionary models of \citet{Siess2000}.}
 \label{tbl_ysos}
 \begin{tabular}{@{ \vline }c@{ \vline }c@{ \vline }c@{ \vline }c@{ \vline }c@{ \vline }}
\cline{1-5}
&&&&\\ 
Region & Source & R.A. &
Decl. & $Age_{JX}$\\
&&(deg)&(deg)&(Myr)\\
(1)&(2)&(3)&(4)&(5)\\
\cline{1-5}
&&&&\\
M17 & 181959.69-161128.1 &   274.998713 &   -16.191166 & 1.9\\
M17 & 182003.75-161339.6 &   275.015630 &   -16.227684 & 3.8\\
M17 & 182007.33-161616.0 &   275.030542 &   -16.271124 & 2.2\\
M17 & 182007.85-161443.1 &   275.032725 &   -16.245325 & 1.2\\
M17 & 182008.08-160938.3 &   275.033700 &   -16.160640 & 0.7\\
M17 & 182010.30-161346.1 &   275.042922 &   -16.229473 & 0.2\\
M17 & 182010.34-161342.7 &   275.043122 &   -16.228539 & 1.1\\
M17 & 182010.92-161036.1 &   275.045500 &   -16.176722 & 0.6\\
M17 & 182011.62-161138.3 &   275.048458 &   -16.193999 & 2.4\\
M17 & 182012.15-161240.7 &   275.050630 &   -16.211330 & 4.7\\
&&&&\\
\cline{1-5} 
\end{tabular}
\end{minipage}
\end{table}

The cluster age gradient analyses reported below in Section \ref{sec_results} are based on the $Age_{JX}$ estimates derived using the Siess00 PMS model. In Section \ref{sec_scales_of_age_gradients} we comment on the change in the size of these gradients assuming other time scales, such as MIST and Fieden16M. 
 
The positions of all young stars with available $Age_{JX}$ estimates (white points) are superimposed on the smoothed stellar surface density maps (Fig.~\ref{fig_maps_individual}). The positions and ages (based on Siess00) for all these stars (1757 MYStIX and 730 SFiNCs) are also listed in Table \ref{tbl_ysos}.  
\begin{table*}\tiny
\centering
 \begin{minipage}{180mm}
 \caption{Age Gradients Within Individual Stellar Clusters. Column 1: A star forming region harboring the cluster of interest. Top 9 and bottom 10 SFRs are from the MYStIX and SFiNCs surveys, respectively. Columns 2-5: Numbers of $Age_{JX}$ YSOs (involved in the age gradient analyses) that lie projected against each of the four annular sub-regions around the cluster's center; with the sub-regions \#1 and \#4 being the closest and farthest from the center, respectively. Columns 6-13: Median cluster age and median radial distance (from cluster's center) for the four annular sub-regions. The age errors are 68\% confidence intervals (CIs) on the medians, estimated using the bootstrap re-sampling technique. Columns 14-15: Median cluster age and median radial distance for the area that combines the sub-regions \#\#2, 3, and 4 (called as sub-region ``2,3,4''). Columns 16-17: Median cluster age and median radial distance for the area that combines the sub-regions \#\#1 and 2 (called as sub-region ``1,2''). Columns 18-19: Median cluster age and median radial distance for the area that combines the sub-regions \#\#3 and 4 (called as sub-region ``3,4''). Columns 20-21. Age differences (in terms of 68\% CIs) between the sub-regions ``2,3,4'' and ``1'', and between the sub-regions ``3,4'' and ``1,2''. For instance, the age difference ``$t_{2,3,4} - t_1$'' is calculated as $(t_{2,3,4} - t_1)/(et_{2,3,4}^{2} + et_{1}^{2})^{1/2}$. Positive (negative) differences indicate increasing (decreasing) age from the cluster's core toward the cluster's periphery.}
 \label{tbl_age_grads_individual}
 \begin{tabular}{@{ \vline }c@{ \vline }c@{ \vline }c@{ \vline }c@{ \vline }c@{ \vline }c@{ \vline }c@{ \vline }c@{ \vline }c@{ \vline }c@{ \vline }c@{ \vline }c@{ \vline }c@{ \vline }c@{ \vline }c@{ \vline }c@{ \vline }c@{ \vline }c@{ \vline }c@{ \vline }c@{ \vline }c@{ \vline }}
\cline{1-21}
&&&&&&&&&&&&&&&&&&&&\\ 
Region & $N_1$ & $N_2$ & $N_3$ & $N_4$ & $t_1$ & $R_1$ & $t_2$ & $R_2$ &  $t_3$ & $R_3$ &  $t_4$ & $R_4$ & $t_{2,3,4}$ & $R_{2,3,4}$ & $t_{1,2}$ & $R_{1,2}$ & $t_{3,4}$ & $R_{3,4}$ & $t_{2,3,4}-t_1$ & $t_{3,4}-t_{1,2}$\\ 
&&&&&(Myr)&(pc)&(Myr)&(pc)&(Myr)&(pc)&(Myr)&(pc)&(Myr)&(pc)&(Myr)&(pc)&(Myr)&(pc)&(68\% CI)&(68\% CI)\\
(1)&(2)&(3)&(4)&(5)&(6)&(7)&(8)&(9)&(10)&(11)&(12)&(13)&(14)&(15)&(16)&(17)&(18)&(19)&(20)&(21)\\
\cline{1-21}
&&&&&&&&&&&&&&&&&&&&\\ 
Eagle &  47 &  47 &  47 &  46 & $ 1.50\pm 0.43$ & 0.41 & $ 2.30\pm 0.37$ & 0.74 & $ 2.10\pm 0.31$ & 1.08 & $ 2.15\pm 0.36$ & 1.49 & $ 2.10\pm 0.19$ & 1.08 & $ 2.10\pm 0.23$ & 0.57 & $ 2.10\pm 0.21$ & 1.26 &  1.28 &  0.00\\
Flame &  31 &  31 &  31 &  29 & $ 0.60\pm 0.17$ & 0.17 & $ 1.20\pm 0.26$ & 0.33 & $ 1.10\pm 0.29$ & 0.53 & $ 1.40\pm 0.28$ & 0.89 & $ 1.20\pm 0.15$ & 0.52 & $ 0.80\pm 0.13$ & 0.24 & $ 1.20\pm 0.18$ & 0.70 &  2.64 &  1.79\\
M 17 & 106 & 106 & 106 & 105 & $ 1.10\pm 0.17$ & 0.52 & $ 1.60\pm 0.23$ & 0.99 & $ 1.40\pm 0.11$ & 1.53 & $ 1.90\pm 0.26$ & 2.73 & $ 1.50\pm 0.13$ & 1.53 & $ 1.30\pm 0.12$ & 0.73 & $ 1.50\pm 0.16$ & 1.93 &  1.93 &  1.01\\
NGC 1893 &  23 &  23 &  23 &  20 & $ 2.70\pm 0.50$ & 0.53 & $ 2.60\pm 0.49$ & 0.94 & $ 1.80\pm 0.36$ & 1.30 & $ 2.90\pm 0.63$ & 1.63 & $ 2.25\pm 0.28$ & 1.28 & $ 2.65\pm 0.36$ & 0.68 & $ 2.20\pm 0.32$ & 1.49 & -0.79 & -0.93\\
NGC 6357 &  30 &  30 &  30 &  29 & $ 1.10\pm 0.19$ & 0.39 & $ 1.90\pm 0.59$ & 0.77 & $ 1.80\pm 0.48$ & 1.40 & $ 1.80\pm 0.82$ & 2.07 & $ 1.80\pm 0.34$ & 1.39 & $ 1.40\pm 0.25$ & 0.59 & $ 1.80\pm 0.42$ & 1.82 &  1.78 &  0.82\\
ONC &  99 &  99 &  99 &  95 & $ 1.50\pm 0.21$ & 0.11 & $ 1.80\pm 0.18$ & 0.33 & $ 1.90\pm 0.23$ & 0.69 & $ 2.00\pm 0.19$ & 0.97 & $ 1.90\pm 0.09$ & 0.68 & $ 1.60\pm 0.13$ & 0.20 & $ 2.00\pm 0.15$ & 0.83 &  1.73 &  1.97\\
RCW 36 &  22 &  22 &  22 &  19 & $ 0.60\pm 0.14$ & 0.14 & $ 1.15\pm 0.13$ & 0.50 & $ 3.00\pm 0.86$ & 0.83 & $ 2.20\pm 0.57$ & 1.48 & $ 1.50\pm 0.25$ & 0.80 & $ 0.90\pm 0.13$ & 0.38 & $ 2.40\pm 0.61$ & 1.09 &  3.16 &  2.40\\
Rosette &  60 &  60 &  60 &  58 & $ 2.80\pm 0.28$ & 0.92 & $ 3.70\pm 0.33$ & 1.76 & $ 3.15\pm 0.45$ & 2.51 & $ 3.35\pm 0.42$ & 3.38 & $ 3.45\pm 0.23$ & 2.48 & $ 3.25\pm 0.29$ & 1.34 & $ 3.30\pm 0.32$ & 2.97 &  1.80 &  0.11\\
W 40 &  26 &  26 &  26 &  24 & $ 0.60\pm 0.08$ & 0.15 & $ 0.80\pm 0.11$ & 0.31 & $ 1.20\pm 0.13$ & 0.60 & $ 1.75\pm 0.29$ & 1.10 & $ 1.15\pm 0.10$ & 0.58 & $ 0.70\pm 0.07$ & 0.25 & $ 1.25\pm 0.14$ & 0.82 &  4.17 &  3.58\\
Be 59 &  31 &  31 &  31 &  29 & $ 2.20\pm 0.45$ & 0.37 & $ 1.50\pm 0.27$ & 0.92 & $ 2.40\pm 0.53$ & 1.50 & $ 2.10\pm 0.48$ & 1.89 & $ 1.90\pm 0.25$ & 1.49 & $ 1.80\pm 0.28$ & 0.59 & $ 2.30\pm 0.33$ & 1.63 & -0.59 &  1.15\\
Cep A &  21 &  21 &  21 &  17 & $ 1.10\pm 0.34$ & 0.43 & $ 2.00\pm 0.35$ & 0.85 & $ 2.10\pm 0.48$ & 1.52 & $ 1.90\pm 0.64$ & 1.86 & $ 2.00\pm 0.23$ & 1.40 & $ 1.65\pm 0.22$ & 0.61 & $ 2.00\pm 0.34$ & 1.65 &  2.19 &  0.87\\
Cep B &  38 &  38 &  38 &  35 & $ 2.10\pm 0.36$ & 0.33 & $ 2.10\pm 0.29$ & 0.62 & $ 2.45\pm 0.62$ & 1.09 & $ 3.50\pm 0.33$ & 1.65 & $ 2.80\pm 0.26$ & 1.07 & $ 2.10\pm 0.21$ & 0.45 & $ 3.00\pm 0.36$ & 1.36 &  1.58 &  2.14\\
Cep C &  10 &  10 &  10 &   7 & $ 0.85\pm 0.33$ & 0.46 & $ 1.35\pm 0.78$ & 0.76 & $ 1.40\pm 0.87$ & 1.27 & $ 4.60\pm 1.10$ & 1.73 & $ 2.00\pm 0.78$ & 1.06 & $ 1.10\pm 0.30$ & 0.65 & $ 2.70\pm 1.04$ & 1.46 &  1.36 &  1.47\\
GGD 12-15 &  12 &  12 &  12 &   9 & $ 1.10\pm 0.53$ & 0.41 & $ 1.30\pm 0.85$ & 0.75 & $ 2.50\pm 0.58$ & 1.07 & $ 3.90\pm 0.91$ & 1.71 & $ 2.50\pm 0.55$ & 1.02 & $ 1.25\pm 0.43$ & 0.58 & $ 2.90\pm 0.61$ & 1.31 &  1.82 &  2.20\\
IC 348 &  30 &  30 &  30 &  27 & $ 2.10\pm 0.23$ & 0.16 & $ 2.65\pm 0.46$ & 0.29 & $ 2.95\pm 0.33$ & 0.52 & $ 3.30\pm 0.44$ & 0.83 & $ 3.00\pm 0.25$ & 0.52 & $ 2.20\pm 0.18$ & 0.21 & $ 3.10\pm 0.30$ & 0.63 &  2.65 &  2.56\\
IC 5146 &  13 &  13 &  13 &  12 & $ 2.00\pm 0.67$ & 0.20 & $ 1.50\pm 0.13$ & 0.43 & $ 2.10\pm 0.50$ & 0.70 & $ 2.50\pm 0.78$ & 1.17 & $ 1.60\pm 0.25$ & 0.68 & $ 1.50\pm 0.18$ & 0.30 & $ 2.10\pm 0.46$ & 0.96 & -0.56 &  1.21\\
LkHa 101 &  21 &  21 &  21 &  17 & $ 1.10\pm 0.33$ & 0.22 & $ 1.60\pm 0.56$ & 0.51 & $ 3.40\pm 0.65$ & 0.75 & $ 2.00\pm 0.48$ & 1.09 & $ 2.20\pm 0.38$ & 0.72 & $ 1.30\pm 0.30$ & 0.32 & $ 2.50\pm 0.56$ & 0.93 &  2.18 &  1.89\\
NGC 7160 &   8 &   8 &   8 &   4 & $ 4.50\pm 0.50$ & 0.55 & $ 3.75\pm 0.59$ & 1.07 & $ 2.75\pm 0.55$ & 1.59 & $ 4.80\pm 0.95$ & 2.27 & $ 3.65\pm 0.59$ & 1.44 & $ 4.10\pm 0.43$ & 0.92 & $ 3.40\pm 0.72$ & 1.82 & -1.11 & -0.83\\
Serpens &   6 &   6 &   6 &   3 & $ 0.45\pm 0.19$ & 0.18 & $ 2.05\pm 1.17$ & 0.29 & $ 2.30\pm 0.70$ & 0.66 & $ 3.90\pm 0.90$ & 1.15 & $ 2.40\pm 0.67$ & 0.54 & $ 0.75\pm 0.40$ & 0.23 & $ 2.40\pm 0.71$ & 0.85 &  2.79 &  2.03\\
&&&&&&&&&&&&&&&&&&&&\\ 
\cline{1-21} 
\end{tabular}
\end{minipage}
\end{table*}

\subsection{Statistical Methods} \label{sec_stat_methods}

All statistical procedures in the current paper were performed using the {\it R} statistical software environment \citep{RCoreTeam2017} including several addon CRAN packages.  Relevant functions include: {\it adaptive.density} in the {\it spatstat} package \citep{Baddeley2015}; {\it ad.test} and {\it qn.test} in the {\it kSamples} package \citep{Scholz2017};  {\it corr.test} in the {\it psych} package \citep{Revelle2017}; {\it cobs} in the {\it cobs} package \citep{Ng2007}; and {\it fisher.test} in the {\it stats} package \citep{RCoreTeam2017}.  

\section{Results} \label{sec_results}

\subsection{Hints Of Age Gradients In Individual Clusters} \label{sec_individual_clusters}

For each of the 19 clusters, their stars are separated into 4 spatially distinct sub-regions requiring that the sub-regions  comprise roughly similar numbers of stars. The $Age_{JX}$ values for individual stars in six exemplifying clusters are shown in Fig.~\ref{fig_age_grads_individual}; similar figure panels for the full sample of 19 clusters are presented in the Supplementary Materials. For each of the 4 sub-regions the median radial distance from the cluster center, the median age and its 68\% confidence interval\footnote{The confidence intervals on medians are obtained using the bootstrap re-sampling technique described in \citet{Getman2014a}. These are roughly equivalent to $\pm 1\sigma$ intervals, but no assumption of Gaussianity is made. No significance levels can be automatically associated with multiples of these intervals; e.g. $p < 0.003$ for $3\sigma$.} are plotted and presented in Table \ref{tbl_age_grads_individual}. These quantities are labeled as $R_1$, $R_2$, $R_3$, $R_4$, $t_1$, $t_2$, $t_3$, and $t_4$. Median distance and age are also calculated for a few blends of various basic sub-regions, such as the sub-regions ``1''+``2'', ``3''+``4'', and ``2''+``3''+``4'', designated as ``1,2'', ``3,4'', and ``2,3,4'', respectively (Table \ref{tbl_age_grads_individual}).  Also listed in the table are the age differences between the sub-regions ``2,3,4'' and ``1'' ($t_{2,3,4}-t_1$) as well as between ``3,4'' and ``1,2'' ($t_{3,4}-t_{1,2}$), normalized by the 68\% confidence interval. The former age difference emphasizes the age gradient between the very core of the cluster and the remaining stars; whereas the later difference underlines possible age gradients on relatively larger spatial scales away from the core.
\begin{figure}
\centering
\includegraphics[angle=0.,width=100mm]{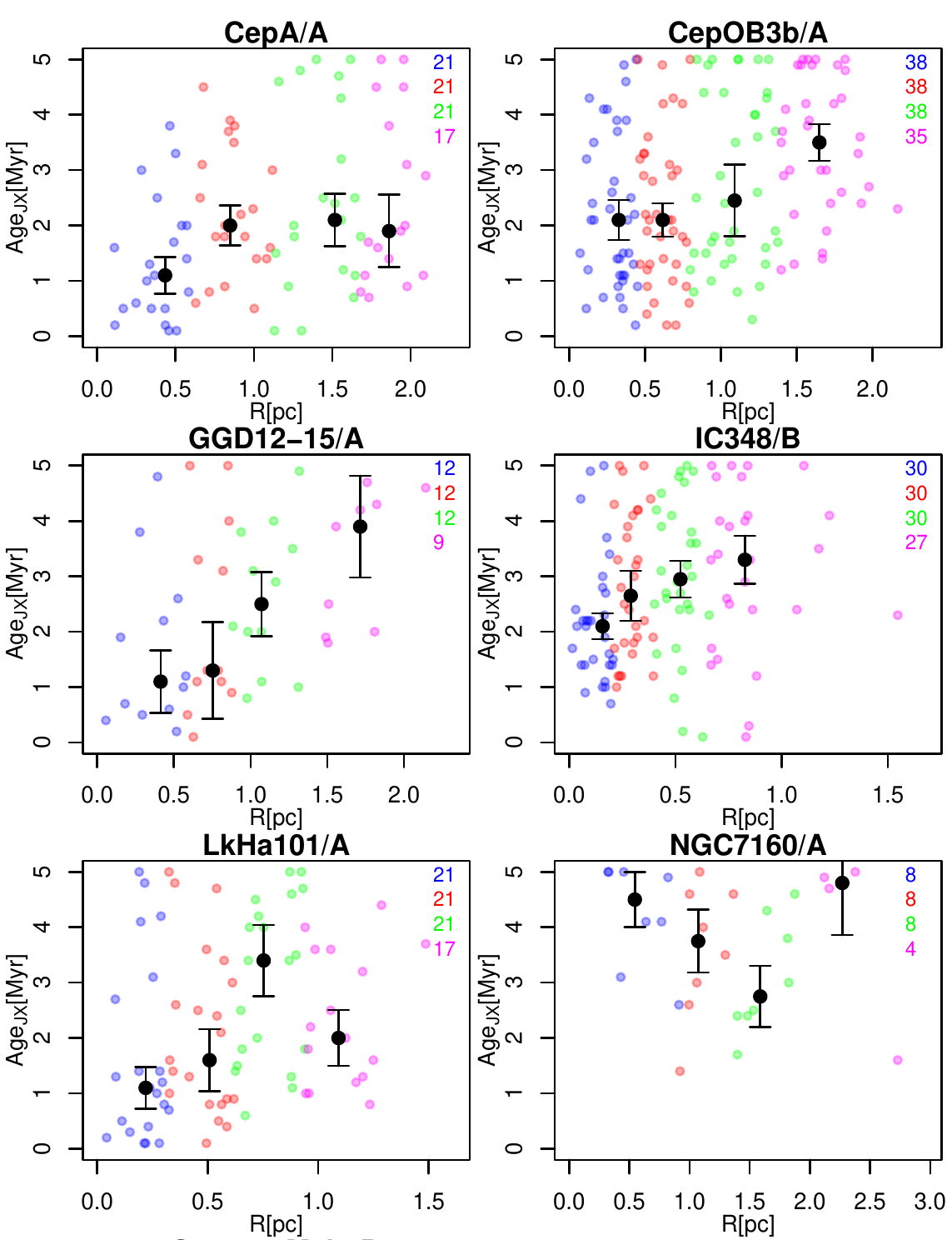}
\caption{Age analysis within individual stellar clusters. Examples are given for six clusters; the complete sample of 19 clusters is shown in the Supplementary Materials. Panels show $Age_{JX}$ as a function of radial distance from the centers of the clusters for individual $Age_{JX}$ stars. Within each cluster, the stars are divided into four spatially distinct annular sub-regions comprising similar numbers of stars (color-coded as blue, red, green, and magenta). For each sub-region, its median($Age_{JX}$) and 68\% confidence interval on the median are marked by a black point and an error bar, respectively. The figure legends provide numbers of plotted stars for each of the four sub-regions. \label{fig_age_grads_individual}}
\end{figure}


The positive (negative) values of $t_{2,3,4}-t_1$ and $t_{3,4}-t_{1,2}$ indicate trends of increasing (decreasing) age from the cluster center toward the cluster periphery. 
\begin{figure*}
\centering
\includegraphics[angle=0.,width=180mm]{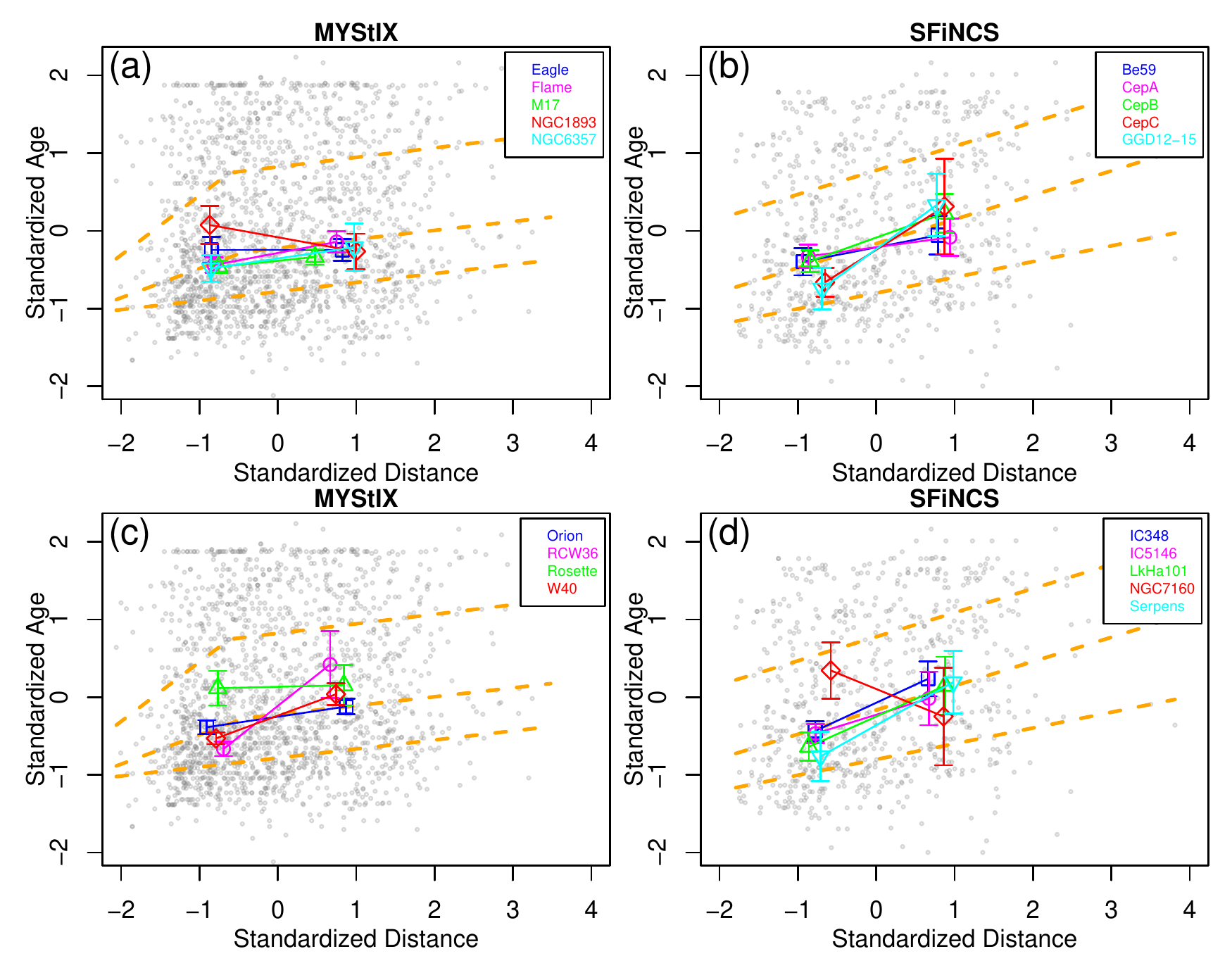}
\caption{Cluster stacking analysis using standardized variables. Standardized age as a function of standardized radial distance from the cluster center for all MYStIX clusters (1757 stars as gray points; left panels) and for all SFiNCs clusters (730 stars as gray points; right panels). The dashed orange lines show the relationship of the 25\%, 50\% (median), and 75\% quartiles of age and distance obtained from B-spline regression. The upper and lower panels have the same points and spline fit, and are reproduced to show clearly the age gradients of individual clusters (colored points).  The colored points on panels (a,b,c,d) indicate the $(t_{3,4} - t_{1,2})$ vs. $(R_{3,4} - R_{1,2})$ age gradients  for individual MYStIX and SFiNCs clusters that were transformed from the physical variables (Table \ref{tbl_age_grads_individual}) to the standardized variables (see the text). \label{fig_standardized_variables_figa}}
\end{figure*}

However, for each of the individual clusters, the statistical significance of the observed age variations is generally low. Only a third of the individual cluster age differences exceed twice the 68\% interval, whereas the remaining age differences are even less significant (Table \ref{tbl_age_grads_individual}).

Multiple different statistical tests, establishing the existence of an age gradient to high statistical significance, are conducted below to ensure that the results are not method dependent. 
\begin{figure}
\centering
\includegraphics[angle=0.,width=90mm]{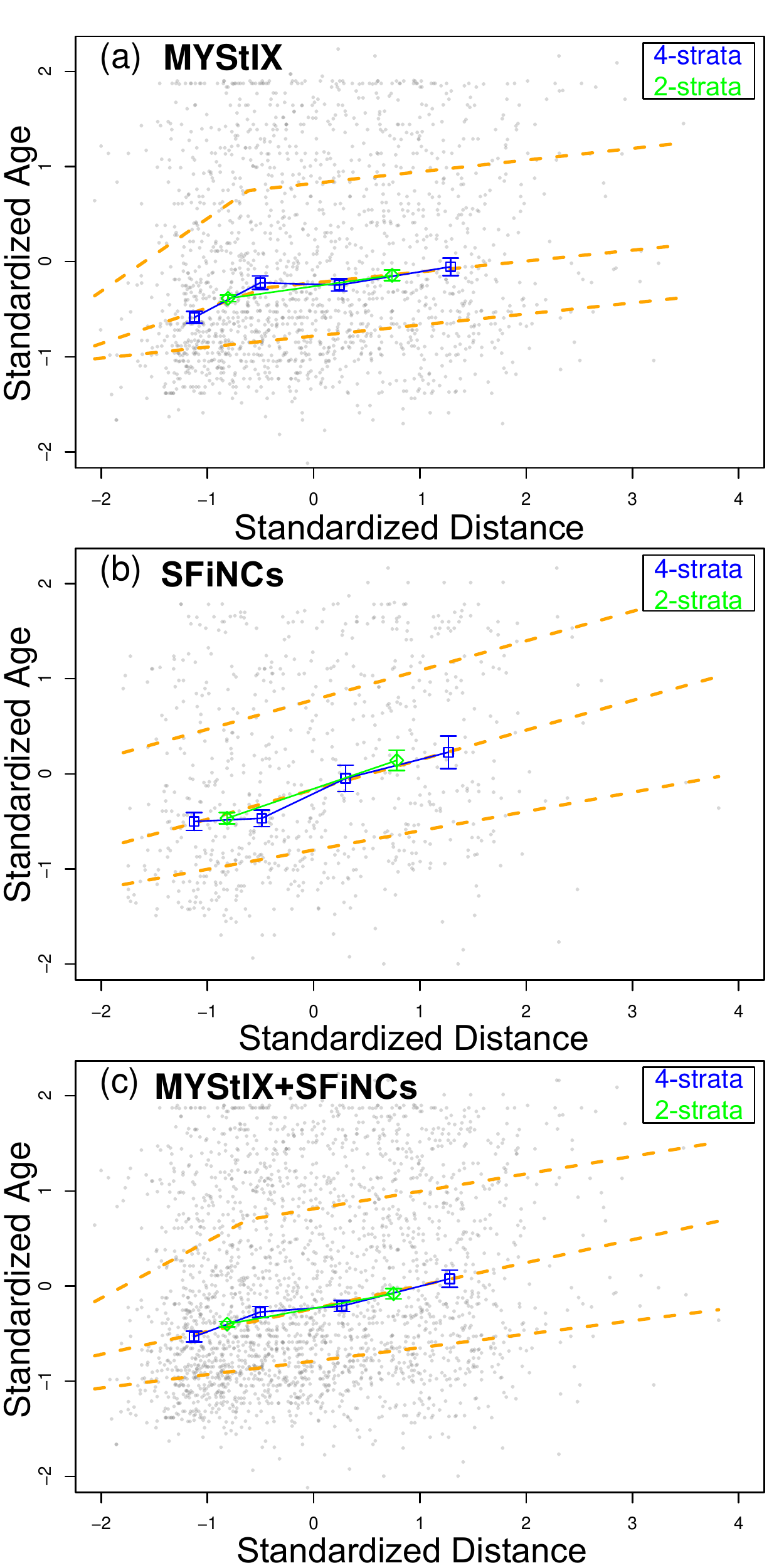}
\caption{Cluster stacking analysis using standardized variables. Standardized age as a function of standardized radial distance from the cluster center for all MYStIX clusters (1757 stars as gray points; panel a), for all SFiNCs clusters (730 stars as gray points; panel b), and for all MYStIX$+$SFiNCs clusters (2487 stars as gray points; panel c). The dashed orange lines show the relationship of the 25\%, 50\% (median), and 75\% quartiles of age and distance obtained from B-spline regression. The colored points on panels (a,b,c) indicate the median age vs. distance values for the merged MYStIX, SFiNCs, and MYStIX$+$SFiNCs cluster samples, respectively. These values are given for the two ``1,2'' and ``3,4'' strata (green) as well as for the four ``1'', ``2'', ``3'', and ``4'' strata (blue). These values are tabulated in Table \ref{tbl_age_grads_combined} as $(t_{s1,2},R_{s1,2})$ and $(t_{s3,4},R_{s3,4})$ (two green strata), and $(t_{s1},R_{s1})$, $(t_{s2},R_{s2})$, $(t_{s3},R_{s3})$ and $(t_{s4},R_{s4})$ (four blue strata). \label{fig_standardized_variables_figb}}
\end{figure}

{\it Statistical Test 1}. For the entire sample of 19 clusters, a random distribution of stellar ages would give a random combination of positive and negative values of $t_{2,3,4}-t_1$ and $t_{3,4}-t_{1,2}$. However, inspection of Table \ref{tbl_age_grads_individual} shows a clear dominance of positive values that indicate older ages in the outer regions of the clusters. Fifteen out of 19 clusters have positive age trends. Bootstrap re-sampling of the distributions of $t_{2,3,4}-t_1$ and $t_{3,4}-t_{1,2}$ show that the observed data are inconsistent with a random age distribution at significance levels of $p<0.009$ and $p<0.0008$, respectively; i.e., cases of negative $t_{2,3,4}-t_1$ and $t_{3,4}-t_{1,2}$ values for half or more of the clusters in a re-sample are very rare. We therefore find that the collective significance level of a positive age trend in the 19 cluster sample has significance $p<0.009$.

\subsection{Age Gradients In Merged Cluster Samples} \label{sec_merged_samples}

We can further investigate the presence of an age gradient by erasing the cluster identifier for each star and considering the stars as age indicators in a merged cluster sample. The clusters are merged together into 3 cluster samples: MYStIX (9 clusters), SFiNCs (10 clusters), and MYStIX+SFINCs (all 19 clusters). We remind the reader that the MYStIX sample has richer OB dominated clusters while the SFiNCs sample has less-rich clusters typically dominated by single late-O or early-B stars. Since the clusters span wide ranges of ages and spatial extends, the merger is performed using standardized age and radial distance variables\footnote{See the meaning of the standardized variable at \url{https://en.wikipedia.org/wiki/Standard_score}.}. In statistics, standardized variables are used for comparison and stacking of observed scores that were measured on different scales. In our case, for each YSO, its standardized variable (age or radial distance from the cluster center) is calculated by subtracting the cluster's mean value from a corresponding individual physical observed variable and then dividing the difference by the cluster's 68\% confidence interval (CI). For the MYStIX, SFiNCs, and MYStIX+SFINCs cluster samples, their standardized ages as functions of standardized radial distances are shown in Figs.~\ref{fig_standardized_variables_figa} and \ref{fig_standardized_variables_figb}. For these merged cluster samples their standardized quantities, analogous to those given in Table \ref{tbl_age_grads_individual}, are presented in Table \ref{tbl_age_grads_combined}.

For the 3 merged cluster samples,  all the three linear regression quartiles\footnote{The quartile linear regression is performed using the R package $cobs$.} (orange lines) in Figs.~\ref{fig_standardized_variables_figa} and \ref{fig_standardized_variables_figb} show clear trends of monotonically increasing age with increasing distance from the cluster center. While the median standard age increases by $\sim 1$ standard unit from the cluster center toward the periphery, the interquartile $[25\% - 75\%]$ range is larger ($\sim 2$ standard units) indicating a wide spread of ages about the median (for instance the MYStIX+SFINCs sample in Fig.~\ref{fig_standardized_variables_figb}).

{\it Statistical Test 2.} Regardless of the large spread, the Kendall's $\tau$ test\footnote{The test is performed using the $corr.test$ R program from the $psych$ package.} yields strong statistically significant positive correlations between the standardized age and distance, with the following $\tau$ coefficient: $\tau = 0.11$ (MYStIX cluster sample), $\tau = 0.17$ (SFiNCs sample), and $\tau = 0.13$ (MYStIX+SFiNCs sample). In all the cases, the correlations are significant at a level $p<<0.0001$.

{\it Statistical Test 3.} The Anderson-Darling and Kruskal-Wallis tests independently support these results\footnote{These tests are performed using the $ad.test$ and $qn.test$ R programs from the $kSamples$ package.}. These tests are used to evaluate the null hypothesis that the standardized age samples corresponding to different cluster sub-regions come from the same underlying age distribution. For each of the three cluster samples (SFiNCs, MYStIX, and SFiNCs+MYStIX) the tests give tiny probabilities (p-values of $p<<0.0001$) that the $t_{s2,3,4}$ and $t_{s1}$ distributions come from a common population (same results are obtained when $t_{s3,4}$ is compared to $t_{s1,2}$).

{\it Statistical Test 4.} The inferred inner ($t_{s2,3,4}-t_{s1}$) and outer ($t_{s3,4}-t_{s1,2}$) age differences correspond to $>3.2 \times$CI and $>5.4 \times$CI for the MYStIX/SFiNCs and MYStIX+SFiNCs cluster samples, respectively (Table~\ref{tbl_age_grads_combined}, Fig.~\ref{fig_standardized_variables_figb}). The Mood's median test, applied separately to the three cluster samples (SFiNCs, MYStIX, SFiNCs+MYStIX), gives tiny probabilities that the medians of the $t_{s2,3,4}$ versus $t_{s1}$ distributions (or $t_{s3,4}$ versus $t_{s1,2}$ distributions) are identical. The resulting p-values are: $p<0.002$ in the case when $t_{s2,3,4}$ is compared to $t_{s1}$ for the SFiNCs sample; and $p<<0.0001$ for the rest of the cases.

Thus the merged cluster samples show the highly statistically significant age gradients, with stellar ages increasing from the cluster center toward the periphery.

\subsection{Age Gradients In Physical Units} \label{sec_scales_of_age_gradients}

The analyses in Sections \ref{sec_individual_clusters} and \ref{sec_merged_samples} establish the existence of an age gradient to high statistical significance in several ways. We can now quantify the size of this gradient in physical units of Myr~pc$^{-1}$.

\begin{figure}
\centering
\includegraphics[angle=0.,width=100mm]{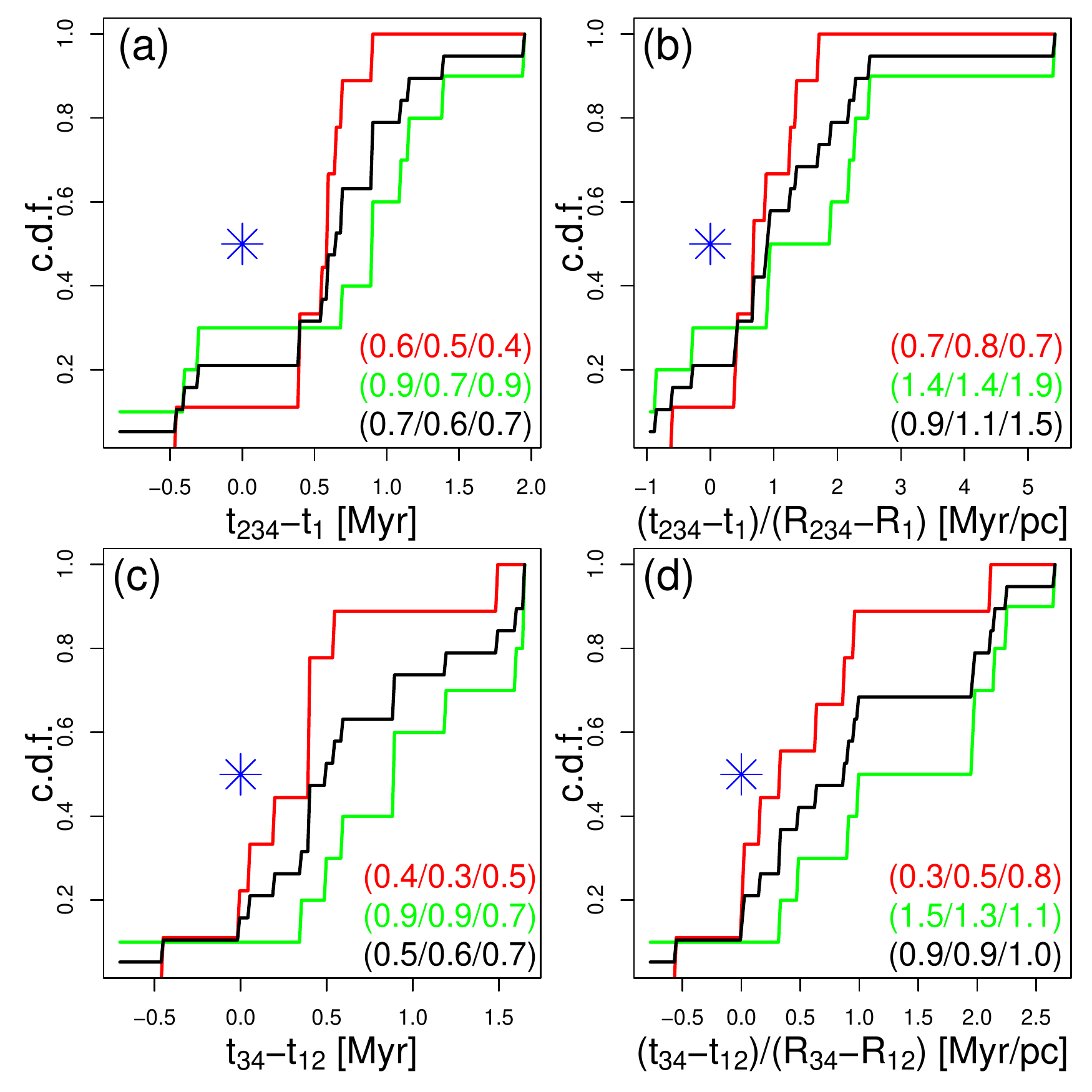}
\caption{Gradient scales in the MYStIX (red), SFiNCs (green), and MYStIX$+$SFiNCs (black) cluster samples. (a) Distributions of the age difference between the ``2,3,4'' and ``1'' annular sub-regions (see Table \ref{tbl_age_grads_individual} for data details). (b) Distributions of the age gradient between the ``2,3,4'' and ``1'' annular sub-regions. (c) Distributions of the age difference between the ``3,4'' and ``1,2'' annular sub-regions. (d) Distributions of the age gradient between the ``3,4'' and ``1,2'' annular sub-regions. The figure legends provide information on the median, mean, and standard deviation of the distributions, in the form of ``(median/mean/stdev)''. Had the MYStIX and SFiNCs data complied with the null hypothesis of having randomly positive and negative age gradients, their c.d.f. curves would have passed through the blue star point, at (0.0, 0.5). \label{fig_gradient_scales}}
\end{figure}

The cumulative distribution functions (c.d.fs.) of the age and the related age gradients are compared between the MYStIX (red) and SFiNCs (green) cluster samples in Fig.~\ref{fig_gradient_scales}. Such distributions are also presented for the merged MYStIX+SFiNCs sample (black). In agreement with the results of the previous analyses, the c.d.fs do not pass through the point at (0.0,0.5) indicating that they do not comply with the null hypothesis of having randomly positive and negative age gradients. 

Panels (b) and (d) suggest that the SFiNCs clusters tend to have higher age gradient scales (medians of $1.4-1.5$~Myr~pc$^{-1}$) than the MYStIX clusters (medians of $0.3-0.7$~Myr~pc$^{-1}$).  However, this difference is not significant according to an Anderson-Darling test. For the MYStIX versus SFiNCs datasets shown in panels (a), (b), (c), and (d) the p-values are 0.16, 0.18, 0.06, and 0.11, respectively.

Considering that the aforementioned MYStIX and SFiNCs distributions do not differ significantly, it is reasonable to combine them together (as MYStIX+SFiNCs; black) to estimate characteristic age gradient scales in our clusters. Figure panels (b) and (d) show that both types of age gradients have similar age gradients of $0.9$~Myr~pc$^{-1}$ (black curves). This age gradient value is based on the Siess00 time scale.
\clearpage
\newpage
\begin{landscape}
\begin{table*}\tiny
\centering
 \begin{minipage}{180mm}
 \caption{Age Gradients For The Merged Cluster Samples (Standardized Variables). Unlike in Table \ref{tbl_age_grads_individual}, here the age and radial distance from the cluster center are standardized variables; thus these are dimensionless quantities (see the text). Column 1: Merged cluster sample. ``SFiNCs'', ``MYStIX'', and ``MYStIX+SFiNCs'' comprise 10, 9, and 19 clusters, respectively. Columns 2-5: Numbers of YSOs (involved in the age gradient analyses) within each of the four annular sub-regions around the cluster's center; with the sub-regions \#1 and \#4 being the closest and farthest from the center, respectively. Columns 6-13: Median cluster age and median radial distance (from cluster's center) for the four annular sub-regions. The age errors are 68\% confidence intervals on the medians, estimated using the bootstrap re-sampling technique. Columns 14-15: Median cluster age and median radial distance for the area that combines the sub-regions \#\#2, 3, and 4 (called as sub-region ``2,3,4''). Columns 16-17: Median cluster age and median radial distance for the area that combines the sub-regions \#\#1 and 2 (called as sub-region ``1,2''). Columns 18-19: Median cluster age and median radial distance for the area that combines the sub-regions \#\#3 and 4 (called as sub-region ``3,4''). Columns 20-21. Age differences (in terms of 68\% confidence interval) between the sub-regions ``2,3,4'' and ``1'', and between the sub-regions ``3,4'' and ``1,2''. For instance, the age difference ``$ts_{2,3,4} - ts_1$'' is calculated as $(ts_{2,3,4} - ts_1)/(ets_{2,3,4}^{2} + ets_{1}^{2})^{1/2}$. Positive (negative) differences indicate increasing (decreasing) age from the cluster's core toward the cluster's periphery.}
 \label{tbl_age_grads_combined}
 \begin{tabular}{@{ \vline }c@{ \vline }c@{ \vline }c@{ \vline }c@{ \vline }c@{ \vline }c@{ \vline }c@{ \vline }c@{ \vline }c@{ \vline }c@{ \vline }c@{ \vline }c@{ \vline }c@{ \vline }c@{ \vline }c@{ \vline }c@{ \vline }c@{ \vline }c@{ \vline }c@{ \vline }c@{ \vline }c@{ \vline }}
\cline{1-21}
&&&&&&&&&&&&&&&&&&&&\\ 
Sample & $N_1$ & $N_2$ & $N_3$ & $N_4$ & $ts_1$ & $Rs_1$ & $ts_2$ & $Rs_2$ &  $ts_3$ & $Rs_3$ &  $ts_4$ & $Rs_4$ & $ts_{2,3,4}$ & $Rs_{2,3,4}$ & $ts_{1,2}$ & $Rs_{1,2}$ & $ts_{3,4}$ & $Rs_{3,4}$ & $ts_{2,3,4}-ts_1$ & $ts_{3,4}-ts_{1,2}$\\ 
&&&&&&&&&&&&&&&&&&&(68\% CI)&(68\% CI)\\
(1)&(2)&(3)&(4)&(5)&(6)&(7)&(8)&(9)&(10)&(11)&(12)&(13)&(14)&(15)&(16)&(17)&(18)&(19)&(20)&(21)\\
\cline{1-21}
&&&&&&&&&&&&&&&&&&&&\\ 
SFiNCs &  183 &  183 &  183 &  181 & $-0.50\pm 0.10$ & -1.13 & $-0.47\pm 0.10$ & -0.49 & $-0.05\pm 0.14$ &  0.30 & $ 0.23\pm 0.16$ &  1.27 & $-0.08\pm 0.08$ &  0.30 & $-0.47\pm 0.06$ & -0.82 & $ 0.14\pm 0.10$ &  0.78 &  3.28 &  5.15\\
&&&&&&&&&&&&&&&&&&&&\\
\cline{1-21}
&&&&&&&&&&&&&&&&&&&&\\
MYStIX &  440 &  440 &  440 &  437 & $-0.59\pm 0.06$ & -1.12 & $-0.22\pm 0.07$ & -0.50 & $-0.25\pm 0.06$ &  0.24 & $-0.05\pm 0.09$ &  1.29 & $-0.19\pm 0.04$ &  0.24 & $-0.39\pm 0.03$ & -0.81 & $-0.14\pm 0.06$ &  0.74 &  5.34 &  3.78\\
&&&&&&&&&&&&&&&&&&&&\\
\cline{1-21}
&&&&&&&&&&&&&&&&&&&&\\
MYStIX &&&&&&&&&&&&&&&&&&&&\\
$+$ &  622 &  622 &  622 &  621 & $-0.53\pm 0.05$ & -1.13 & $-0.27\pm 0.06$ & -0.50 & $-0.21\pm 0.06$ &  0.27 & $ 0.08\pm 0.09$ &  1.28 & $-0.14\pm 0.04$ &  0.26 & $-0.40\pm 0.03$ & -0.81 & $-0.08\pm 0.05$ &  0.75 &  5.74 &  5.44\\
SFiNCs &&&&&&&&&&&&&&&&&&&&\\
&&&&&&&&&&&&&&&&&&&&\\ 
\cline{1-21} 
\end{tabular}
\end{minipage}
\end{table*}
\end{landscape}
\clearpage
\newpage

According to the time scale transformations given in Section \ref{sec_ages_only}, the age gradients based on the MIST and Feiden16M time scales are expected to be by 17\% lower and 67\% higher than the ones based on Siess00. So that the typical age gradients are $\sim 0.75$~Myr~pc$^{-1}$ and $1.5$~Myr~pc$^{-1}$ assuming the MIST and Feiden16M time scales, respectively.

\section{Discussion} \label{sec_discussion}

Several years ago, \citet{Getman2014b} discovered that stars in the cluster cores of two nearby young clusters (ONC and Flame) are systematically younger (that is, formed later) than stars in the cluster periphery. In this study, we demonstrate that this property is generally present in young, rich, isolated clusters.  Eighty percent of the MYStIX and SFiNCs clusters examined here show stellar ages increasing with radial distance from the cluster center. While often weak in individual clusters, the collective trend is highly statistically significant. The median amplitude of the age gradients is $\sim 0.75$~Myr~pc$^{-1}$, 0.9~Myr~pc$^{-1}$, and 1.5~Myr~pc$^{-1}$ when defined on the time scales of MIST, Siess00, and Feiden16M, respectively.

Other researchers have reported similar age gradients in star forming regions (Serpens North, Serpens South, Corona Australis, W 3 Main, and IRAS 19343+2026; references in \citet{Getman2014b}). More recently, an independent study of the ONC cluster based on the optical OmegaCAM survey by \citet{Beccari2017} reports the detection of three distinct episodes of star formation. In agreement with our original discovery of the core-halo age gradient in ONC (based on X-ray and infrared datasets of young stars), the optical data of Beccari et al. present evidence for spatio-age gradients in ONC with the younger stellar populations lying projected in a tight concentration within the cluster core and the oldest stellar population more sparsely distributed in and around the core. These results are reminiscent of the spatio-age stellar distributions in ONC previously found in the HST-based study by \citet{Reggiani2011}. Beccari et al. speculate that the star formation in ONC might have progressed along the line of sight with the youngest population lying further away from the observer. But this explanation requires a particular geometry and can not account for the presence of age gradients in most young star clusters.

\citet{Getman2014b} provide several plausible scenarios that alone, or in combination with each other, may lead to the observed core-halo age gradients in young stellar clusters. We briefly review these here.

Scenario A is related to the density thresholds in star formation, which would allow the star formation to continue in the cluster core but to cease in the less dense halo. Scenario B invokes the global gravitational contraction and star formation threshold that would lead to the acceleration of star formation rate and formation of younger stars in the denser cloud/cluster core. Scenario C relies on the condition of the late formation of massive stars in the core, accompanied by the formation of some low-mass siblings. Scenario D is based on the kinematic stellar drift in a supervirial cluster and/or stellar ejections by three-body interactions, which would produce an older halo. Scenario E considers the dynamical stellar heating during cluster relaxation as the main source of older halo. Scenarios F and H invoke the merger and expansion of multiple sub-clusters with older sub-cluster populations largely contributing to the halo of the resultant cluster. Scenario G considers the fall of gas filaments into a potential well and further replenishment of the cluster core with younger generations of stars.

Among theoretical astrophysical calculations of star cluster formation, the most promising in explaining the observed age gradients, is the global hierarchical collapse (GHC) model by \citet*{Colin2013} and \citet*{VS2017} that leads to a hierarchical assembly of a stellar cluster exhibiting core-halo age gradients. Some of the qualitative star formation scenarios from \citet{Getman2014b}, such as B, C, D, F, H, and G, are coherently included in this quantitative model. The GHC model describes a molecular cloud in a free-fall state that globally gravitationally contracts (with a little contribution from turbulence or magnetic field) to form filaments. Relatively small stellar groups form at different times in different locations of the filaments. These stellar groups and gas continue falling towards the center of the gravitational potential, mixing and replenishing the gas and star material in the central molecular clump that is forming a young stellar cluster. Massive stars tend to form later within the central clump, at the peak of the star formation rate. The feedback from massive stars cuts off the flow of material from the in-falling filaments leading to the destruction of the molecular clump and the cessation of star formation in the cluster. Older stars, which formed in multiple in-falling stellar groups prior to reaching the cluster center, tend to exhibit large velocity dispersions and thus appear further away from the cluster center at the end of star formation, producing core-halo age gradients.

Similar behaviors of in-falling, star forming filaments are seen in the simulations of \citet{Bate2009,Bate2012}; but these simulations cover only $10^5$ years, which is too short to investigate longer timescale age gradients seen from the observations of young stellar clusters.

Judging from Fig.~9b of \citet{VS2017}, the stars in the halo ($r>0.6$~pc) of their modeled cluster Group-12 have a median age of $\sim 2$~Myr at a median distance from the cluster center of $r \sim 1.2$~pc, whereas the stars in the core ($r<0.6$~pc) have a median age of $\sim 1$~Myr at a median distance of $r \sim 0.2$~pc; this results in an age gradient of $\sim 1$~Myr~pc$^{-1}$. This agrees well with the gradient of 0.9~Myr~pc$^{-1}$ (based on Siess00) observed in the MYStIX+SFiNCs clusters.

The inferred MYStIX+SFiNCs age gradient of $0.9$~Myr~pc$^{-1}$ (corresponding to a speed of 0.9~km~s$^{-1}$) also agrees well with the stellar velocity dispersions of $\sim 1$~km~s$^{-1}$ recently observed in a number of young, nearby, embedded stellar clusters, such as NGC~1333, L~1688, and Chamaeleon~I \citep{Foster2015,Rigliaco2016,Sacco2017}.  Thus, stars drifting away from their site of formation in a dense cloud with a velocity dispersion of $\sim 1$~km~s$^{-1}$ could also produce the observed core-halo age gradients (our Scenario D above). 

The late formation of massive stars in cluster cores has independent evidence. Radial velocity measurements suggest that the 30~M$_\odot$ star $\theta^1$C Ori that dominates the ONC Trapezium core is only $\sim 10^4$ years old \citep{ODell2009}, and dynamical calculations support this very young age for the Trapezium core \citep{Allen2017}.  The embedded cluster W 3 Main has a core with both extremely young and older massive stars \citep{Tieftrunk1998} surrounded by a larger cluster of older stars \citep{Feigelson2008}. Popular scenarios for the formation of massive stars include the quasi-static monolithic collapse model \citep{McKee2002} and the dynamical scenarios of either stellar mergers \citep{Bonnell2002} or accretion from gas streams, such as the ones present in the aforementioned GHC model \citep{VS2017, Motte2017}. The outcomes of both dynamical scenarios, GHC and stellar mergers, are consistent with the observed late formation of massive stars. According to the latter scenario, some younger more massive stars may have formed through collisions of older less massive stars during the relatively short high-density phase of a cluster core as modelled by \citet{Bonnell2002}. Recall, however, that the age gradient reported here and by \citet{Getman2014b} refers only to lower mass stars (roughly $0.2-1.2$~M$_{\odot}$; based on Siess00 scale) rather than massive stars.

The empirical finding reported in the present study -- late or continuing formation of stars in the cores of star clusters with older stars dispersed in the outer regions -- thus has a strong foundation with other observational studies and with the astrophysical models like the global hierarchical collapse model of \citet{VS2017}.

Our study refers to young stellar clusters; all but two (Rosette/D and NGC 7160/A) analyzed here clusters have median ages below $3$~Myr (based on the Siess00 scale). No positive age trends are detected in NGC 7160/A. The recent {\it Chandra} study of the rich, isolated NGC 6231 cluster \citep{Kuhn2017}, with the median age above $3$~Myr (based on Siess00), find no a spatial gradient in ages. It is thus possible that such older clusters either ``never had a positive radial age gradient or that a previously existing gradient disappeared with age as a result of dynamical mixing'' \citep{Kuhn2017}.

\section{Conclusions} \label{sec_summary}

The current paper reports the discovery of the core-halo age gradients, with younger cores and older halos, in the vast majority of the morphologically simple, isolated, and rich MYStIX and SFiNCs clusters. Eighty percent of the analyzed clusters show upward core-halo age trends (Section \ref{sec_individual_clusters}). The collective effect of these trends is highly statistically significant (Section \ref{sec_merged_samples}). Depending on the choice of PMS evolutionary model, the inferred median age gradient within the MYStIX+SFiNCs clusters varies from $0.75$~Myr~pc$^{-1}$ (based on MIST) to $0.9$~Myr~pc$^{-1}$ (based on Siess00) to $1.5$~Myr~pc$^{-1}$ (based on Feiden16M) (Section \ref{sec_scales_of_age_gradients}). The observed spatio-age gradients can be explained within the framework of the sophisticated astrophysical calculations of star cluster formation such as the global hierarchical collapse model (Section \ref{sec_discussion}).

\section*{Acknowledgements}

We thank the referee, C. R. O'Dell, for helpful comments. The MYStIX project is now supported by the {\it Chandra} archive grant AR7-18002X. The SFiNCs project is supported at Penn State by NASA grant NNX15AF42G, {\it Chandra} GO grant SAO AR5-16001X, {\it Chandra} GO grant GO2-13012X, {\it Chandra} GO grant GO3-14004X, {\it Chandra} GO grant GO4-15013X, and the {\it Chandra} ACIS Team contract SV474018 (G. Garmire \& L. Townsley, Principal Investigators), issued by the {\it Chandra} X-ray Center, which is operated by the Smithsonian Astrophysical Observatory for and on behalf of NASA under contract NAS8-03060. The Guaranteed Time Observations (GTO) data used here were selected by the ACIS Instrument Principal Investigator, Gordon P. Garmire, of the Huntingdon Institute for X-ray Astronomy, LLC, which is under contract to the Smithsonian Astrophysical Observatory; Contract SV2-82024. This research has made use of NASA's Astrophysics Data System Bibliographic Services and SAOImage DS9 software developed by Smithsonian Astrophysical Observatory.












\bsp	
\label{lastpage}
\end{document}